\documentclass{eas}
\usepackage{graphicx}
\begin{document}

\title{Dynamical architectures of planetary systems induced by orbital 
migration}
\runningtitle{Architectures of planetary systems} 
\author{E. Szuszkiewicz}
\address{CASA* and Institute of Physics, University of Szczecin, 
Wielkopolska 15, 70-451 Poland}
\author{J. C. B. Papaloizou}\address{DAMTP, University of Cambridge, 
Wilberforce Road, Cambridge CB3 0WA, UK}
\begin{abstract}
The aim of this talk is to present the most recent advances 
in establishing plausible planetary system architectures 
determined by the gravitational tidal interactions between the planets and 
the disc in which they are embedded during the early epoch of planetary 
system formation. We concentrate on a very well defined and intensively 
studied process of the disc-planet interaction leading to the planet 
migration. We focus on the dynamics of the systems in which 
low-mass planets are present. Particular attention is devoted 
to
investigation of the role 
of resonant configurations. Our studies, 
apart from being complementary to the fast progress occurring just now 
in observing the whole variety of planetary systems and uncovering their 
structure and origin, can also constitute 
a valuable contribution 
in support
of the missions planned to enhance
the number of detected multiple 
systems.
\end{abstract}
\maketitle
\section{Diversity of planetary systems and migration-induced resonances}
Multiple-planet systems are of particular interest to test theoretical
models of planet formation and the evolution of planetary systems. 
Just a few days before this conference, Jason Wright (Wright {\em et al.\/} 
\cite{wright}) sent 
a paper in which he and his colleagues
generated a catalogue of the 27 published multiple-planet systems
around stars within 200 pc from the Sun 
to the other participants.
The authors
 made several interesting 
observations about the systems' statistical properties. At least five of these
systems harbour planets which are in or near mean-motion resonance. This
fact makes 
the importance of orbital migration even
stronger than immediately after the discovery of giant planets very close 
to their host stars (e.g. Kley \cite{kley}; Papaloizou {\em et al.\/} 
\cite{fivebig}). These arguments, together
with the increasing number of known planets with masses in the range of 
a few Earth masses, have stimulated our study of migration-induced
resonances (e.g. Papaloizou \& Szuszkiewicz \cite{ps}, Podlewska \&
Szuszkiewicz \cite{edytas}) in systems containing low-mass planets.
 
Our present goal is to investigate migration-induced architectures
of planetary systems, focusing mainly on resonant configurations.
In order to achieve this goal we have performed a series of numerical
simulations of planets migrating in resonance. The main tools used in our
studies are two-dimensional hydrodynamic simulations, simple analytic 
modelling and N-body investigations. By constructing a simple analytic model,
we were able to verify the reliability of our numerical calculations. 
By combining the hydrodynamic simulations
with the N-body technique we 
were able to follow the dynamical evolution 
of the planets for a substantial amount of time, comparable with the estimated
lifetimes of gaseous discs.   

\section{Initial configuration and computational set-up}
\begin{figure}
\centering
\centerline{\includegraphics[width=10cm]{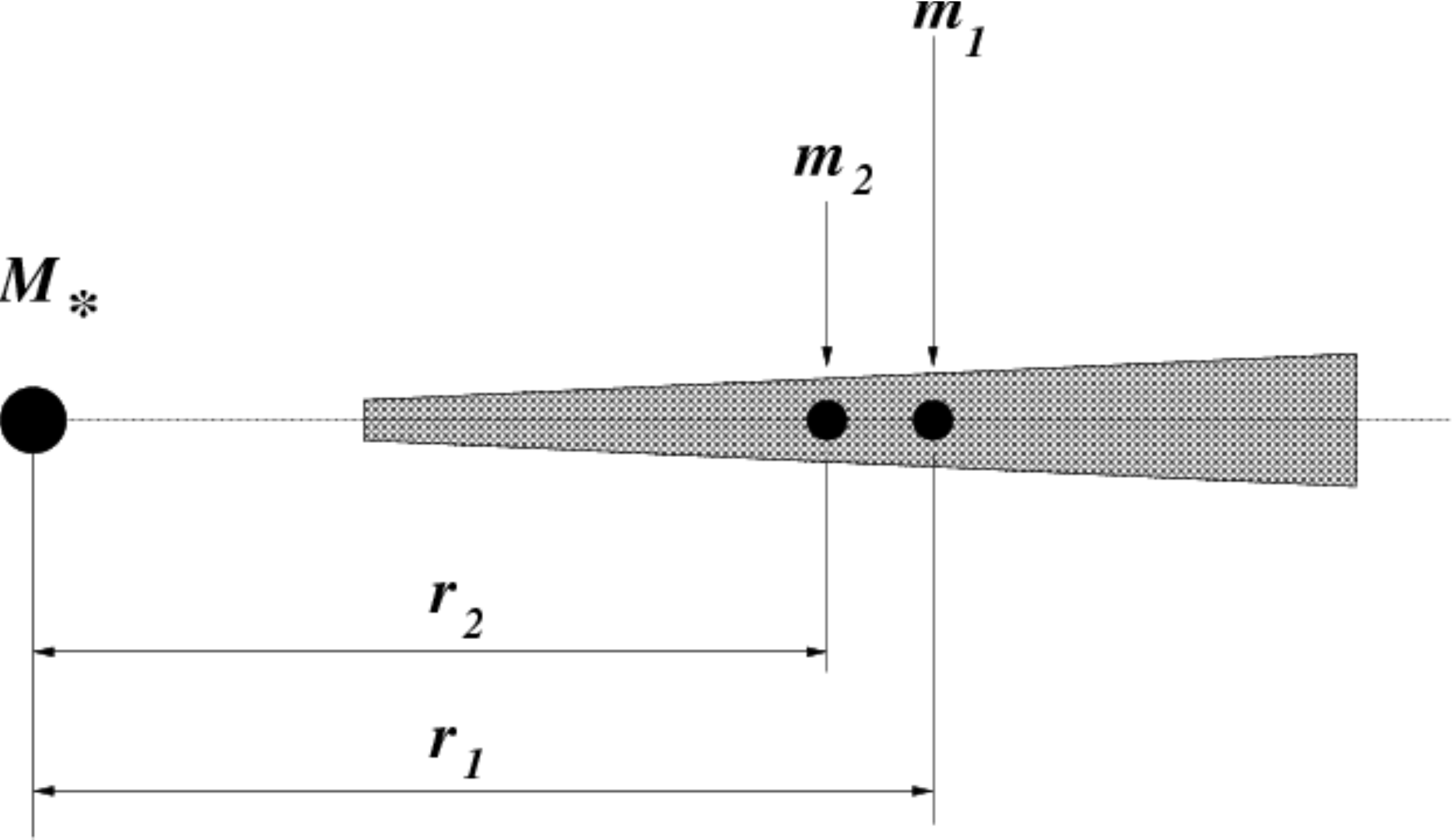}}
\caption{\label{fig1}{The initial configuration: two planets, with
masses $m_{1}$ and $m_{2}$ respectively, in circular orbits around
a central star with mass $M_*$ at distances $r_{1}$ and $r_{2}$,
are embedded in a gaseous disc.
}}
\end{figure}

Our initial set-up shown in Fig.~\ref{fig1} includes the central
star with a mass $M_*$ and two orbiting planets with masses $m_{1}$
and $m_{2}$ respectively. The two planets
are embedded in a disc which is the source of planet orbit migration.
They are initialised on circular orbits around
a central mass  which has a fixed value of 1 solar mass.
The gravitational potential was softened with softening parameter
$b =0.8H$, where $H$ is the semi-thickness of the disc. 
This results in the formation of an equilibrium atmosphere
around the embedded planet which then does not accrete.
This softening also allows an adequate representation of type
I migration in two dimensional discs
(see e.g. Nelson \& Papaloizou \cite{nelpap}).
The disc in which planets are initially  embedded has a
surface density
profile $\Sigma(r)$  essentially flat (see 
Papaloizou \& Szuszkiewicz \cite{ps} for a full description). 
The planets are located in the flat part of this distribution.
We use four different values for
the maximum value of the surface density $\Sigma_0$,
namely the standard value 
of
$2\times 10^3$ kg/m$^2$,
attributed to the
minimum mass solar nebula at 5.2AU 
 which we denote $\Sigma_1$,
then $\Sigma_{0.5}=0.5 \Sigma_1$, $\Sigma_{2}= 2 \Sigma_1$ and
finally  $\Sigma_{4}= 4\Sigma_1$.
The radial boundaries were taken to be open.

The calculations presented here were performed with a two-di\-men\-sio\-nal
version of the Eulerian hydrodynamic code NIRVANA. For details of the
numerical scheme and code adopted, see Nelson {\em et al.\/} 
(\cite{nelson2000}).

\section{A Super-Earth ($m_1 = 4M_{\oplus}$) and an Earth analogue 
($m_2=1M_{\oplus}$)}
\label{tanakas}
Let us consider here a system of two planets with disparate masses, namely
a Super-Earth with a mass of 4M$_{\oplus}$ and an Earth analogue 
(1M$_{\oplus}$).
The time-scale  of inward migration for a low-mass planet on a circular 
orbit embedded in  a disc with constant surface density can be approximated
by the formula given by Tanaka {\em et al.\/}  (\cite{tanaka})
\begin{equation}
\tau_r = \left|{r_p\over {\dot r_p}}\right |
=W_m{M_* \over m_{p}} {M_* \over \Sigma_p r_p^2}
\left({c \over r_p \Omega_p}\right)^2 \Omega_p^{-1} 
\label{MIG}
\end{equation}
Here $M_*$ is the mass of the central star, $m_{p}$ is the mass
of the planet orbiting  at distance $r = r_p$, $\Sigma _p$ is the
disc  surface density at $r = r_p$, $c$ is
the local sound speed and $\Omega_p$ is the
angular velocity at $r = r_p$.
The numerical coefficient $W_m$ is $0.3704$.

The density waves excited by a low mass planet with small eccentricity
in the disc  lead to orbital circularization (e.g. Artymowicz \cite{ar93}; 
Papaloizou \& Larwood \cite{pl00})
at a rate that can be estimated as (Tanaka \& Ward \cite{tw04})
\begin{equation}
t_c = {\tau_{r}\over W_c} \left({c\over r_p \Omega_p}\right)^2.
\label{CIRC}
\end{equation}
Here, the numerical coefficient $W_c$ is equal to $0.289$.

\begin{figure*}
\begin{minipage}{125mm}
\centering
\hbox{
\hbox{\includegraphics[width=6.0cm]{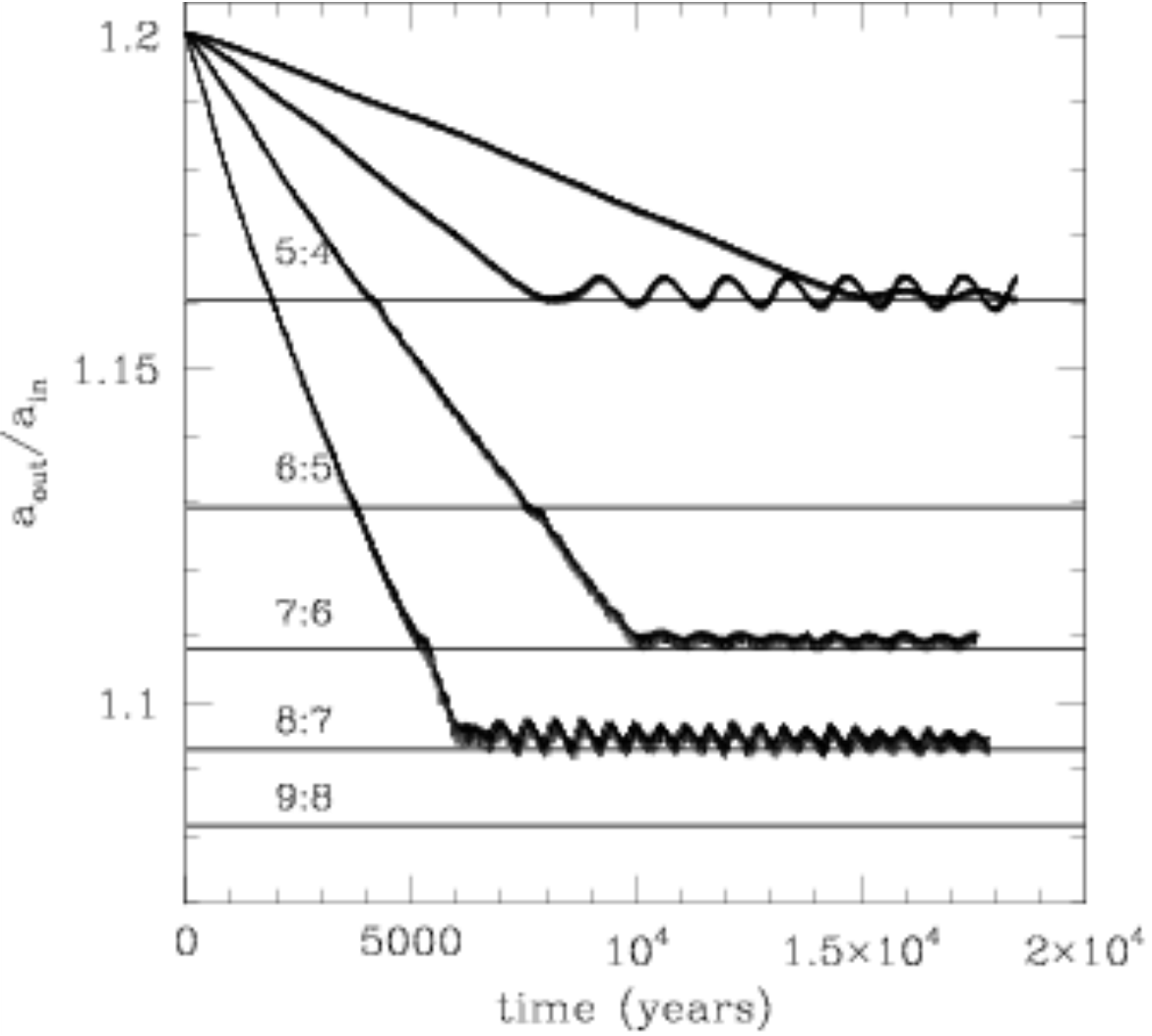}}
\hbox{\includegraphics[width=6.0cm]{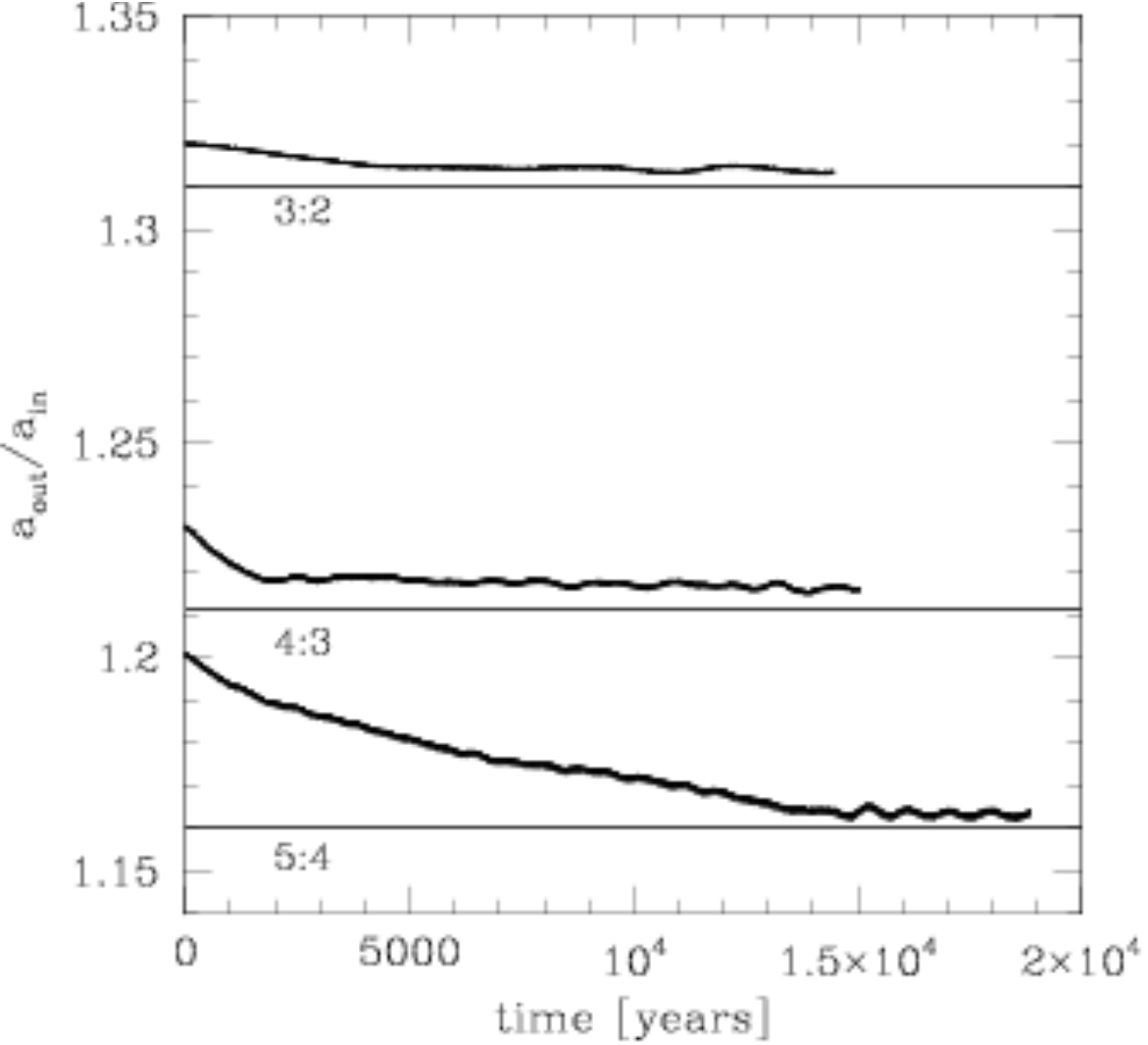}}
}
\caption{\label{fig2}{({\em left}) The evolution of the ratio of semi-major
axes for the two planets with masses $m_1= 4M_{\oplus}$ and
$m_2 = 1M_{\oplus}.$ Starting from the lower curve and going
upwards, the curves correspond to the  initial surface
density scalings  $\Sigma_0 = \Sigma_{4}$,
$\Sigma_0 =\Sigma_{2}$, $\Sigma_0 =\Sigma_{1}$, and
$\Sigma_0 =\Sigma_{0.5}$ respectively.
({\em right})
The semi-major axis ratio when
$m_1 = m_2 = 4M_{\oplus}$ with
$\Sigma_0 = \Sigma_{1}$ (uppermost curve) and with
$\Sigma_0 = \Sigma_{4}$ (two lower curves) }}
\end{minipage}
\end{figure*}

It is expected  from equation (\ref{MIG}) that
two planets with different masses will migrate at different rates. In our 
case, the
Super-Earth in the external orbit will migrate faster than the Earth analogue.
This has the consequence that their period ratio will evolve
with time and may accordingly attain and become  locked in a mean-motion 
resonance (Nelson \& Papaloizou \cite{nelpap02}; Kley {\em et al.\/} 
\cite{kley04}).
We have considered  the pair
of planets to be evolving in discs with  $ \Sigma_0 = \Sigma_{0.5}$,
$ \Sigma_0 = \Sigma_{1}$, $\Sigma_0 = \Sigma_{2}$ and
$\Sigma_0 = \Sigma_{4}.$ The evolution was  followed
until a resonance  was established. The results are summarised
in Fig.~\ref{fig2} (left panel) where the evolution of the ratio of the
semi-major axes, which starts from the value 1.2 for all four
cases, is shown.
The fastest migration (steepest
slope) corresponds to the  case   where
the two planets
are embedded in a disc with $\Sigma_0 = \Sigma_{4}$ and the slowest
to the case of a disc with $\Sigma_0 = \Sigma_{0.5}$. The planets
in the disc with $\Sigma = \Sigma_{4}$  become
trapped in a $p+1:p$ = 8:7 resonance. If the disc surface density is two times
smaller, then a 7:6 resonance is attained. If it is four or eight
times smaller, then the attained resonance is 5:4.
These results are fully consistent with the idea that higher
$p$ resonances are associated with faster relative migration
rates. Let us concentrate on the 8:7 resonance, illustrating our
combined methodology of  hydrodynamic calculations  (an explicit
evolution of a 2D gaseous disc with embedded planets is calculated), 
simple analytic modelling and 
N-body techniques (the migration and eccentricity damping due to
the gaseous disc is included in the form of analytic prescriptions).

\subsection{Two-dimensional hydrodynamic simulations}
\begin{figure*}
\centering
\vbox{
\hbox{ \hbox{\includegraphics[width=4.3cm,angle=270]{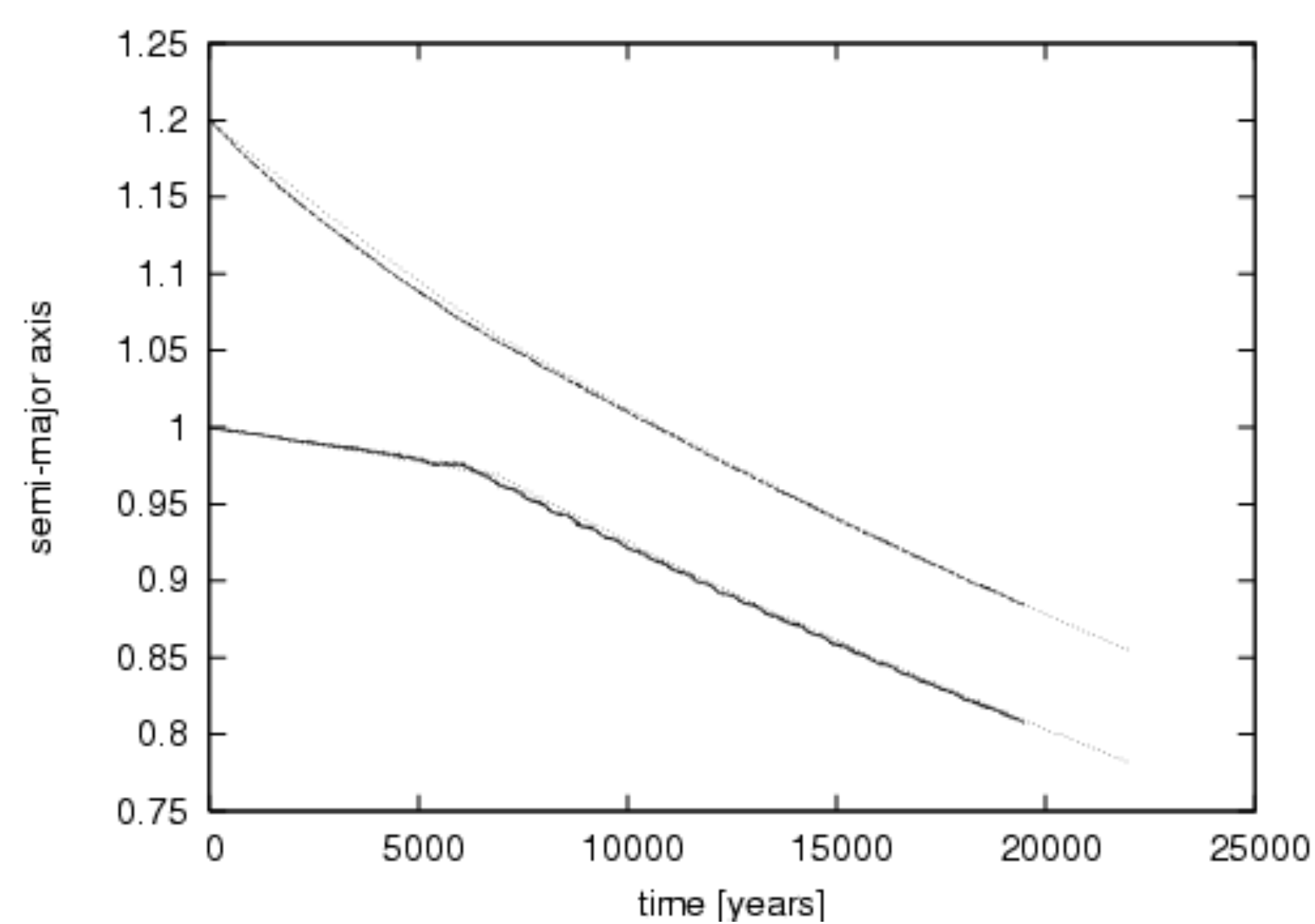}}
       \hbox{\includegraphics[width=4.3cm,angle=270]{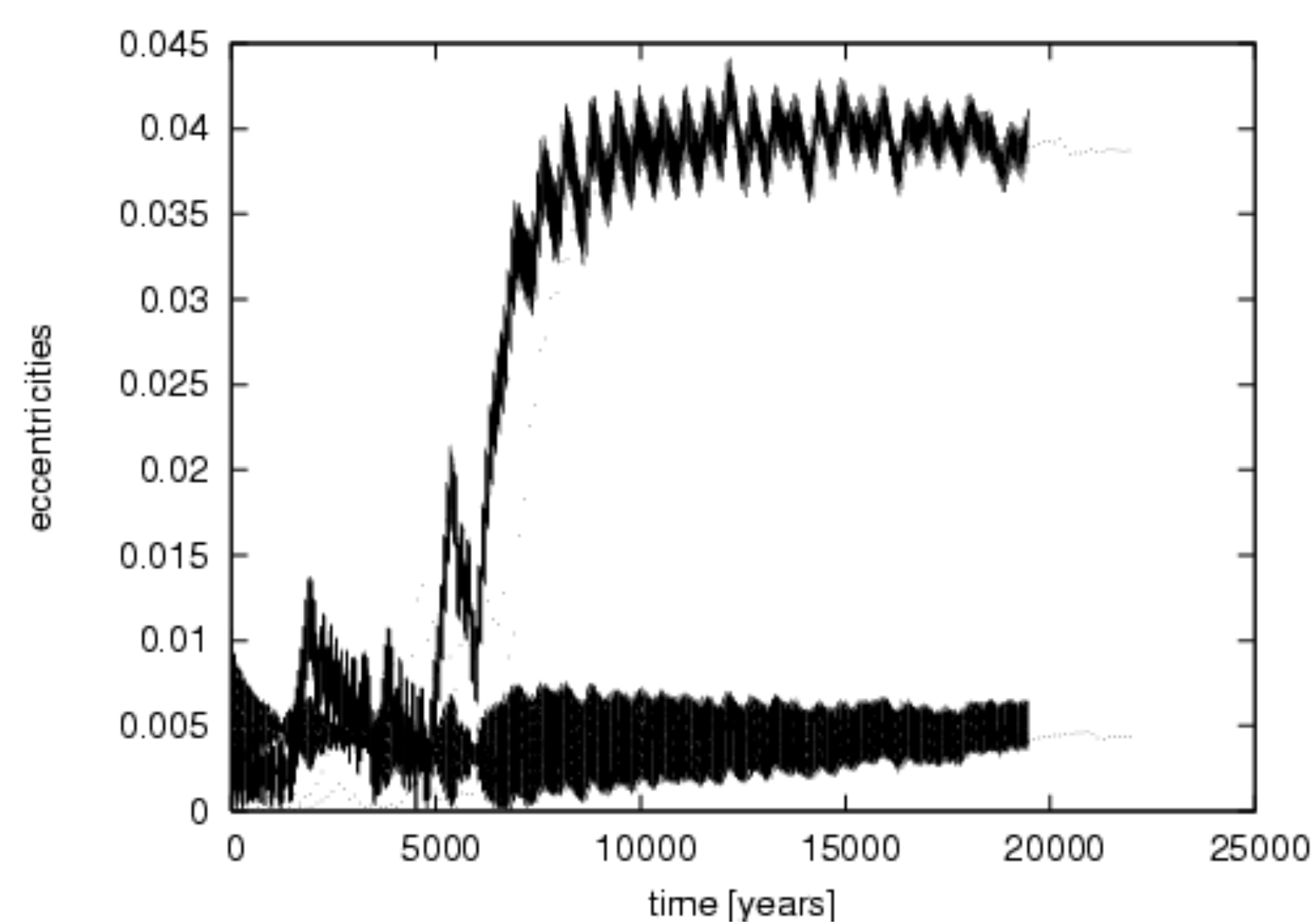}}
     }
\hbox{ \hbox{\includegraphics[width=4.3cm,angle=270]{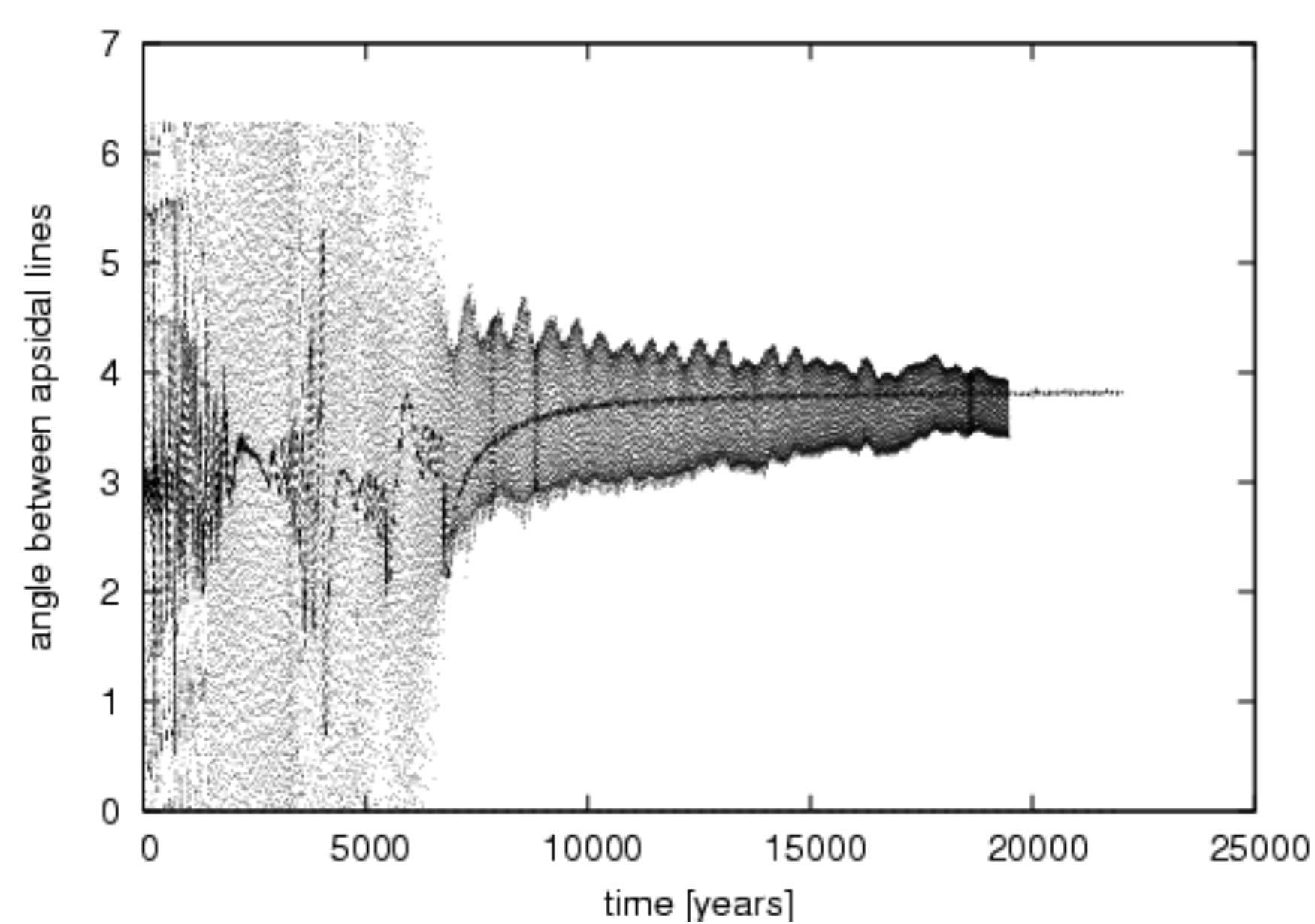}}
       \hbox{\includegraphics[width=4.3cm,angle=270]{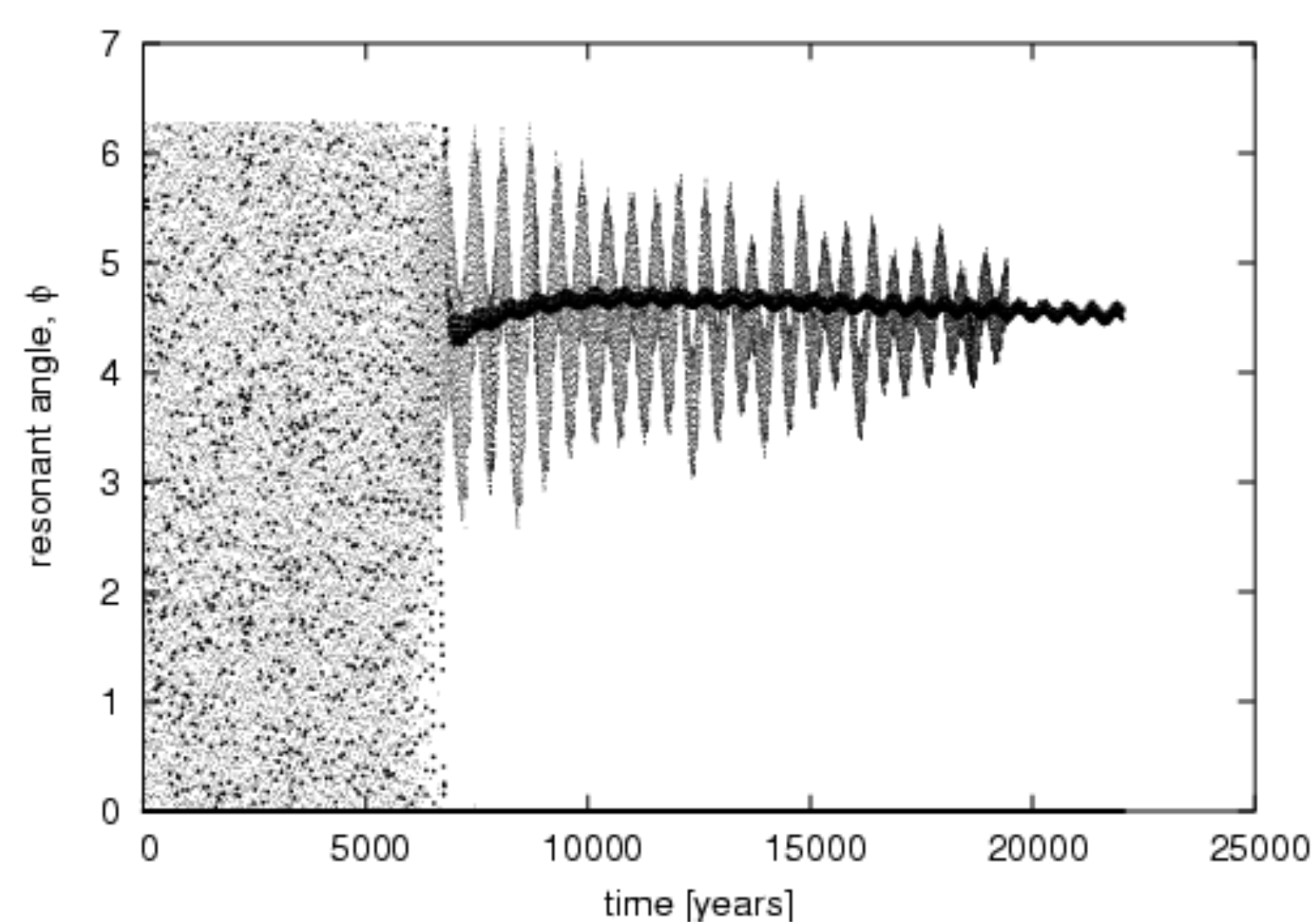}}
     }
}
\caption{\label{fig3}{The evolution of semi-major axes, eccentricities,
angle between apsidal lines and resonant angle
for two planets with masses, $m_{1} = 4M_{\oplus}$ and
$m_{2} = 1M_{\oplus}$ migrating towards
a central star embedded in the disc with 
$\Sigma =\Sigma _4$,
obtained by hydrodynamic simulations and     
N-body simulations (dotted lines in two upper panels
and dark lines in two lower panels).
}}
\end{figure*}

The whole evolution for the fastest migration rate is illustrated
in  Fig.~\ref{fig3}. This shows the evolution
of the individual semi-major axes, eccentricities,
angle between apsidal lines and  one of the resonant angles.
\noindent
The inner planet did not migrate significantly until an 8:7 resonance was
attained at about 6000 years. Subsequently, the two planets migrated
inwards together, maintaining the commensurability. During 18000 years
of evolution the outer planet changed its location from a dimensionless
radius of  1.2 to 0.9. The eccentricity  of the outer planet increased
slightly and that of the inner planet substantially, reaching 
an equilibrium value of 0.04 at
around 10000 years and at  later times
oscillating around this value. At the end of the simulation
shown here, the ratio of eccentricities $e_{1}/e_{2}$ is equal
to 0.125. The local peaks in the values of inner planet eccentricity
occurring at 2000, 4000 and around 5400 years correspond to the
planet passing through the 5:4, 6:5 and 7:6 resonances respectively.
After about 8000 years the angle between the  apsidal lines 
oscillates around 212$^{\circ}$ and the resonant angle $\phi$ around
264$^{\circ}$. The amplitude of the
oscillations for both angles decreases   with time.

\subsection{Comparison with the analytic model}
\label{analytic}
It is of interest to compare the eccentricities obtained above
when the planets are trapped in a commensurability with what
is expected when resonant effects and disc tides are in balance.
Papaloizou \& Szuszkiewicz (\cite{ps}) have given a simple 
approximate analytic solution for two migrating planets locked in
a $p+1:p$ commensurability. In particular, they have shown that
if eccentricities are not too large, then
when they stop growing, 
they must satisfy
\begin{eqnarray}
{e_1^2\over t_{c1}} + {e_2^2\over t_{c2}}{m_2n_1a_1
\over m_1n_2a_2}
 -\left({e_1^2\over t_{c1}}-{e_2^2\over t_{c2}}\right)f
 =  
\left({1\over t_{mig1}}-{1\over t_{mig2}}\right){f\over 3},    
\label{ejcons0}
\end{eqnarray}
where  $ f =m_2a_1/((p+1)(m_2a_1+m_1a_2)).$
Here the semi-major axes and  eccentricities of the two planets
$(i=1,2)$ are $a_i$ and $e_i.$
The migration rates (assumed directed inwards) and circularization times
induced by the disc tides are $t_{migi}  =  |n_i/{\dot n_i}|
=  |2a_i/(3{\dot a_i})| = 2\tau_r/3,$ and
 $t_{ci} =  |e_i/{\dot e_i}|$ respectively.
The mean motions are $n_i$.

In the simulations presented
here $e_2 >> e_1$. We can simplify matters even further, setting
$e_1 =0$. We can also assume that $p$ is large, because in the present example
 its value is 7.
We also note that from equations (\ref{MIG}) and (\ref{CIRC})
it turns out that $t_c = (\tau_{r}/W_c) (H/ r)^2$.
Thus we obtain
\begin{equation}
e_2^2
= \left({m_1 \over m_2}\left({a_2\over a_1 }\right)^{1/2} - 1\right)
{m_1\over 0.578 (p+1)(m_2+m_1)}
 (H/ r)^2
\label{ejcons03}.
\end{equation}
If we apply the above relation to the case illustrated in Fig.~\ref{fig3}
for the two planets with masses $m_{1} = 4M_{\oplus}$ and
$m_{2} = M_{\oplus}$
embedded in a disc with the aspect ratio $H/r =0.05$ and initial surface
density scaling  $\Sigma_0 =\Sigma _4$, we obtain $e_2 =0.037$
in reasonable agreement with the simulations.

\subsection{Comparison between hydrodynamic simulations and N-body
calculations}

The hydrodynamic calculations discussed in the previous sections
allow us to follow the migrations of planets for almost 2$\times$10$^4$
years. As a result, we have been able to simulate planets
becoming trapped in resonances. The behaviour and stability of
the resonance trapping vary with the planet masses  and the
surface density of the disc in which they are embedded.
The outcomes of these simulations could be well matched to those
of  N-body integrations where we  incorporate simple
prescriptions for the migration and eccentricity damping given
through equations  (\ref{MIG}--\ref{CIRC}).
Using these expressions in the N-body code we have extended
the hydrodynamic calculations for a longer period of time
and studied the long term stability of the
resonances that we found. 
As a first step we have adjusted the  numerical coefficients
in equations (\ref{MIG}-\ref{CIRC}) in such a way that the
hydrodynamic  and N-body approaches
give the same qualitative evolution.

As an example, in Fig.~\ref{fig3} we show the results of
the comparison for the case of two planets  with masses
1$M_{\oplus}$ and 4$M_{\oplus}$ respectively embedded
in a  disc with initial  surface density scaling parameter
$\Sigma_0 = \Sigma_4$.
The numerical coefficients adopted were   $W_m = 0.3647$
and $W_c = 0.225$.
These results show good agreement with the hydrodynamic
results and display the same 8:7 commensurability.
The fitted coefficients were also  reasonably close to those expected
from analytic disc-planet interaction theory.

\begin{figure*}
\centering
\vbox{
\hbox{ \hbox{\includegraphics[width=4.3cm,angle=270]{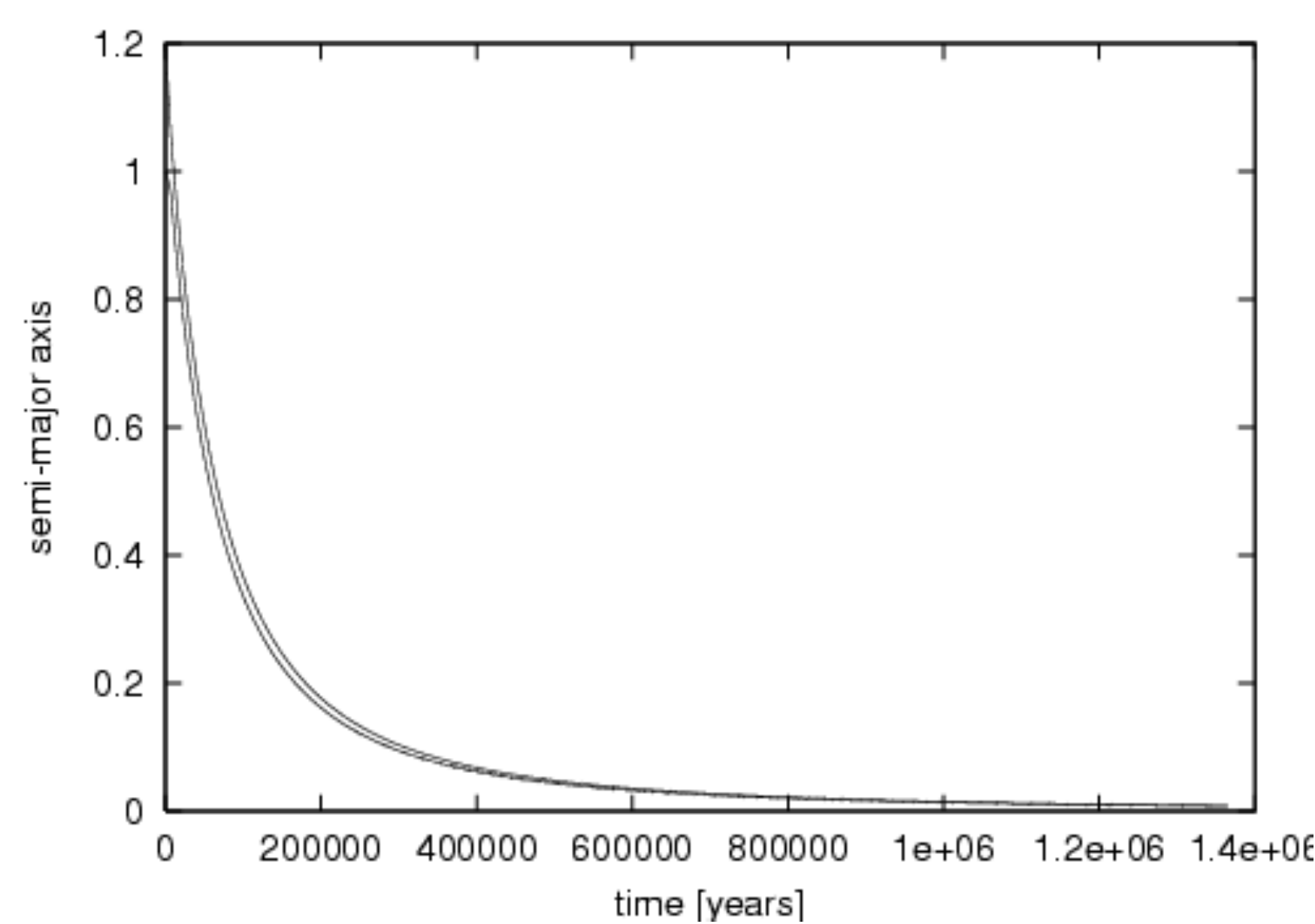}}
       \hbox{\includegraphics[width=4.3cm,angle=270]{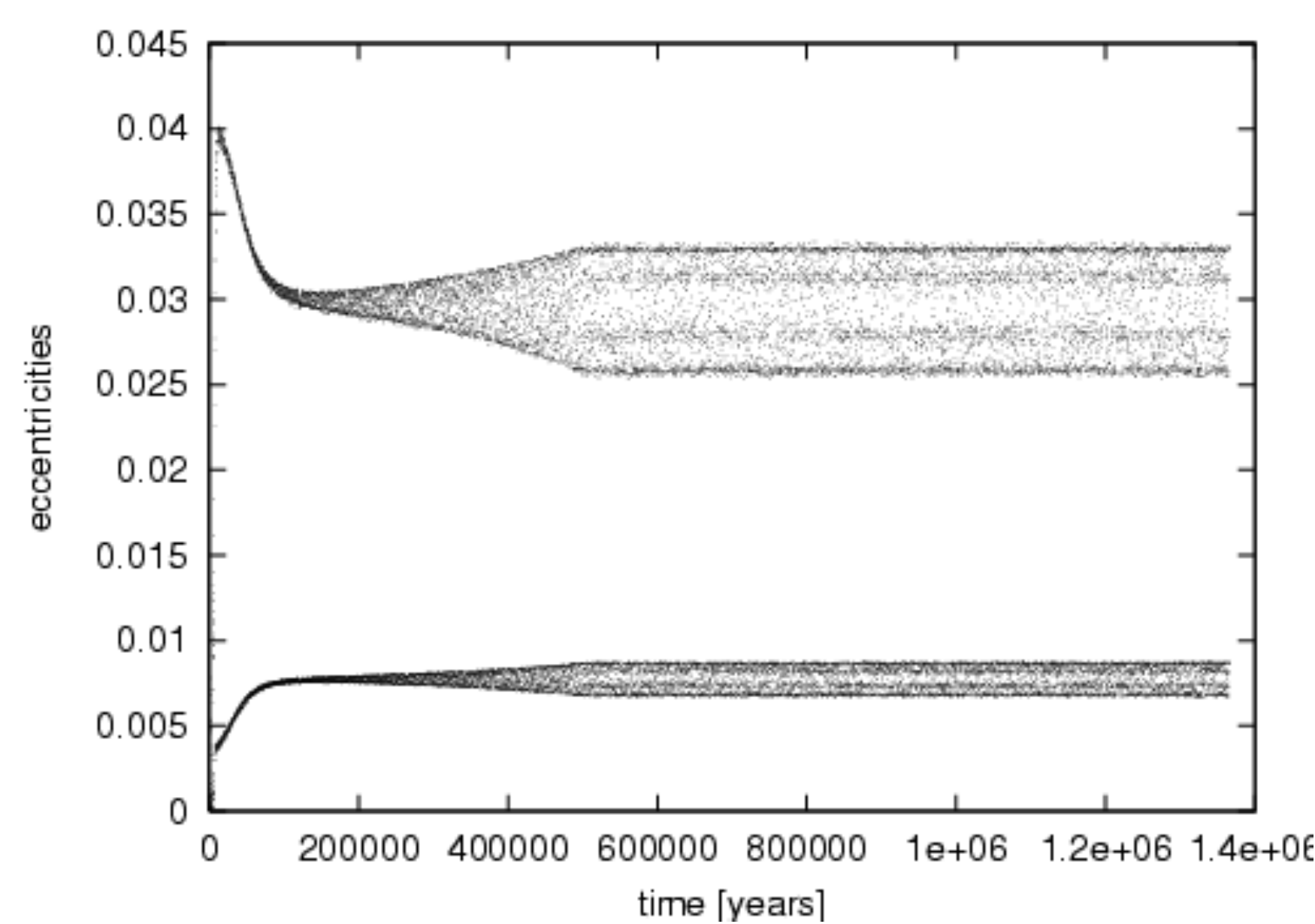}}
     }
}
\caption{\label{fig4}{The evolution of the  semi-major axes and  eccentricities
for two planets with masses $m_{1} = 4M_{\oplus}$ and
$m_{2} = 1M_{\oplus}$ migrating towards
a central star embedded in the disc with $\Sigma =\Sigma _4$
obtained from N-body simulations.
}}
\end{figure*}

The evolution was followed using the N-body approach for
an additional 1.2~$\times 10^{6}$ years.
The two planets remain in the 8:7 resonance during this
time and there is no indication of any significant change
in the  monitored quantities for the last 9~$\times 10^{5}$
years. The long term evolution of this
system is shown in Fig.~\ref{fig4}. It is interesting
to note that the equilibrium value of eccentricity of the
inner planet is around $e_2=0.03$, in accordance with the
value predicted by the simple analytic
model discussed in Section \ref{analytic}.

\section{Two Super-Earths ($m_1 = m_2 = 4M_{\oplus}$)}
The resonant interaction must balance
the tendency towards  relative migration of the two planets.
This is smaller when the planets have the same mass.  In 
order to investigate this fact,
two planets of  mass $4M_{\oplus}$ were initiated  close to a
3:2 resonance  with $r_{1}=1.32$ and $r_{2}=1$ in a disc
with  $\Sigma_0 = \Sigma_{1}$.
The  evolution of
the  semi-major axis ratio for the planet  orbits  is shown
in Fig.~\ref{fig2} (right panel).
We also performed  simulations
in a disc
with  $\Sigma_0 = \Sigma_{4}$
 for the same masses  starting
at $r_{1}=1.23$ and $r_{2}=1.00$  and at $r_{1}=1.20$, $r_{2}=1.00$.
The planets   become trapped in the   nearest available
resonance which is a good indication of stability.
The   evolution of the equal mass pair of planets is discussed
more extensively in Papaloizou \& Szuszkiewicz (\cite{ps}) 
and in {\L}acny \& Szuszkiewicz (this volume).

\section{A gas giant and a Super-Earth ($m_1 = 333M_{\oplus}$, 
$m_2 = 5.5M_{\oplus}$)}
\begin{figure*}
\centering
\vbox{
\hspace*{-5mm}\hbox{ \hbox{\includegraphics[width=5.8cm]{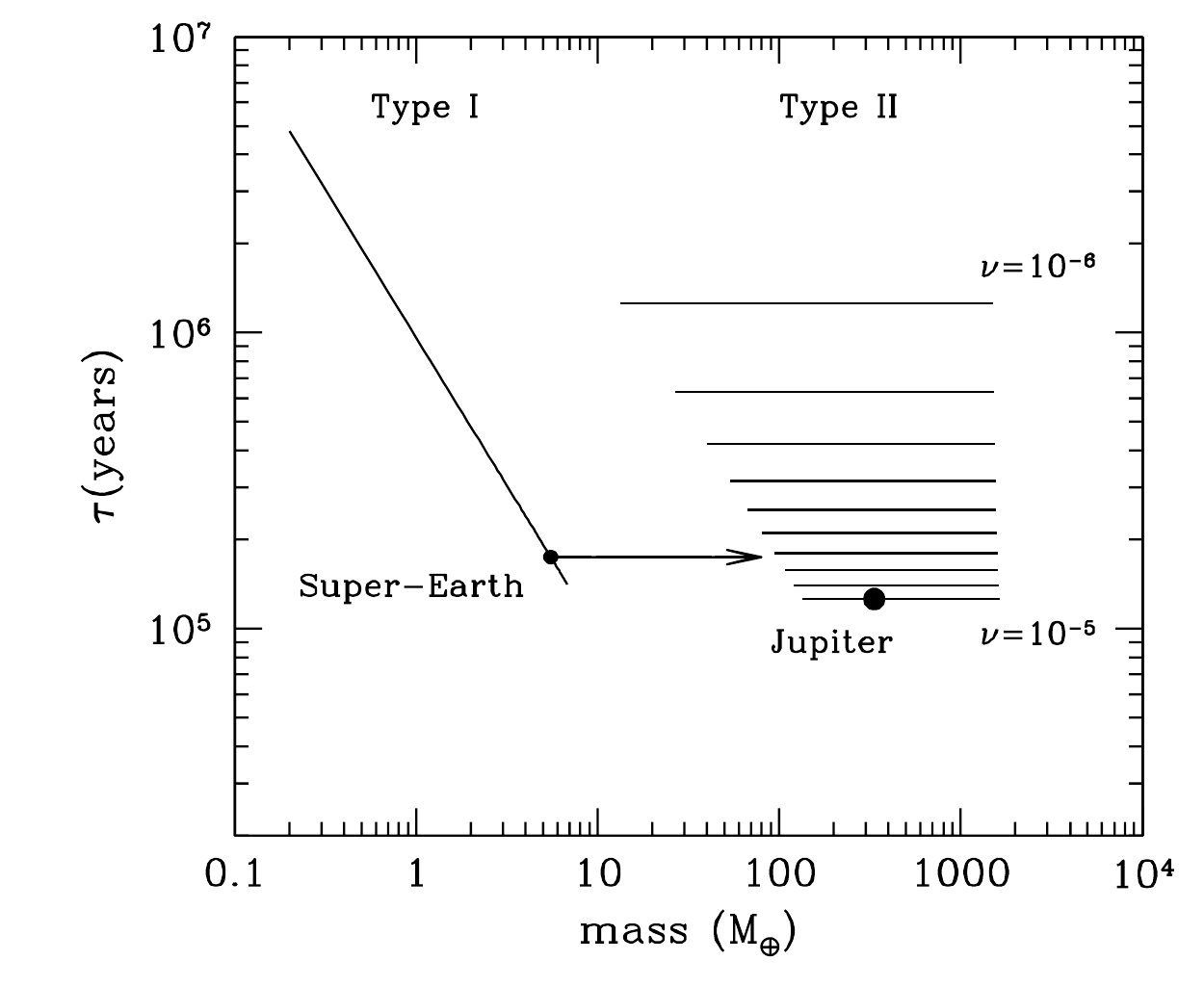}}
       \hbox{\includegraphics[width=6.6cm]{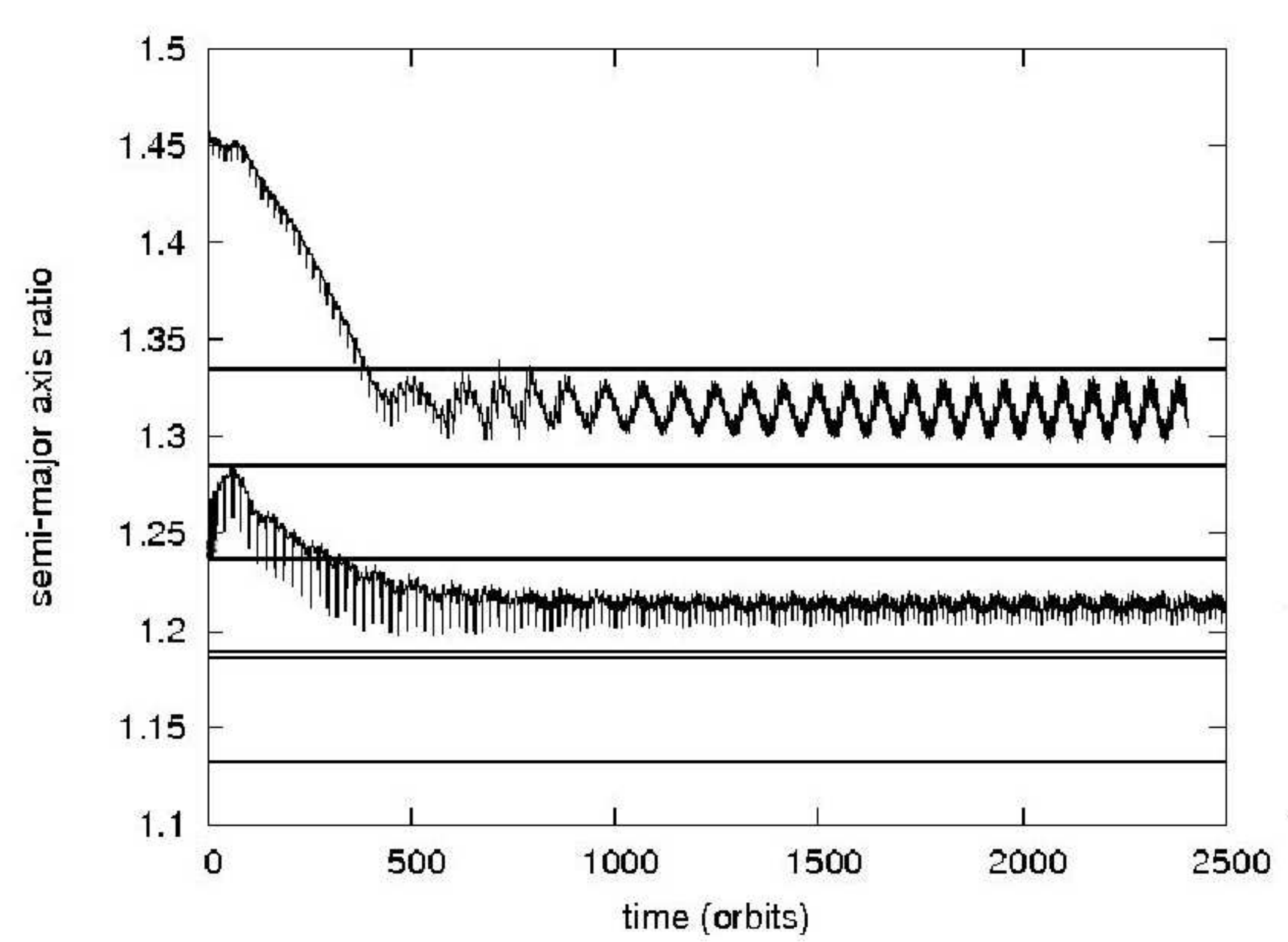}}
     }
}
\caption{\label{fig5}{({\em left}) Comparison between
the times of migration for planets with different masses located  at 5.2 AU,
embedded in a disc with a flat surface density distribution, the
standard value attributed to the minimum mass solar nebula at 5.2 AU, namely 
$\Sigma$ = 2000 kg/m$^2$  and  H/r = 0.05.
({\it right})
The evolution of the ratio of semi-major axes for the Jupiter and Super-Earth
embedded in a gaseous disc. In the case of the upper curve
the initial semi-major axis ratio is 1.45 and planets became locked into
a 3:2 resonance. For the lower curve, the initial semi-major axis ratio is
1.24 and the attained commensurability is 4:3. The solid horizontal lines
are the 3:2, 4:3 and 5:4 resonance widths in the case  of circular orbits in 
the restricted three body problem (Wisdom, \cite{wisdom}, 
Lecar {\em et al.\/} \cite{lecar}). 
 }}
\end{figure*}

In the previous sections we have investigated systems in which
both planets have a low mass. Here we extend our study to the case
where one planet is a Super-Earth and the other a gas giant.
The migration rates for different planet masses has been estimated
by a number of authors, see the review by Papaloizou {\em et al.\/}
(\cite{fivebig}).
Their results are illustrated in Fig.~\ref{fig5} (left panel), 
where we plot the
migration time of a planet as a function of its mass. There are two
mass regimes which are of interest here, namely (0.1--30$M_{\oplus}$)
and (150--1500$M_{\oplus}$), for which in a typical protoplanetary disc
we can talk about two different types of migration, called type I
and type II respectively. 
The migration time for low mass planets embedded
in a gaseous disc (type I migration) has been discussed above in
Section~\ref{tanakas}.
Type II migrators open a gap in the disc and their evolution
is determined by
the radial velocity drift in the disc $v_r$.
The migration time can be estimated as follows
 (Lin and Papaloizou \cite{linpap93})
\begin{eqnarray}
\tau_{II}={r_p\over v_r} =\frac{2 {r_p}^2}{3 \nu}
\label{tauii}
\end{eqnarray}
where $\nu$ is a kinematic viscosity parameter.
This has been shown in Fig.~\ref{fig5} (left panel)
for $r_p=5.2 AU$ and different values of $\nu$
(10$^{-5}$,
2 $\times$ 10$^{-5}$, 3 $\times$ 10$^{-5}$, ..., 9 $\times$ 10$^{-5}$ 
and $10^{-6}$) 
expressed in
dimensionless units.
In drawing lines
for a given viscosity parameter
we have taken into account the condition for a gap opening in the disc which
reads
${m_p \over M_*} > {40\nu \over r_p^2 \Omega_p}$.
From this expression, it is clear that the bigger the viscosity  $\nu$,
 the bigger
the mass of the planet must be to be able to open a gap.
We have chosen the mass of the type I and type II migrators to be
5.5$M_{\oplus}$ and one Jupiter mass respectively. Their locations  have been
marked in Fig.~\ref{fig5} according to their masses  and, in the case of
the Jupiter, also according to the viscosity adopted in the disc ($\nu  = 10^{-5}$).
It is obvious from this figure that we should expect a
convergent migration
if the Jovian type planet is in the external orbit
and the Super-Earth in the internal one.
Starting from the configuration illustrated in Fig.~\ref{fig5} (left panel),
we have studied the possible
resonances in this system. 
In order to investigate how the outcome of such evolution depends on the
initial planet configurations,  we have performed our simulations
with a wide range of planetary separations (see Table 1 in
Podlewska \& Szuszkiewicz \cite{edytas}).
The 2:1 commensurability was not taken into account because
of computational time constraints.
The  resonances located closer to the Jupiter, such as
5:4, 6:5 or higher values of $p$, are not possible,
because for such small separations the system becomes unstable
and the Super-Earth is
scattered from the disc.
We have found that two outcomes are
possible, namely either
the Super-Earth is ejected from the disc or the planets
become locked
in 3:2 or 4:3 mean-motion
resonances.
In Fig.~\ref{fig5} (right panel) we show  two examples 
of resonant trappings
occurring when the relative planet separation is 0.45 (upper curve) and
0.24 (lower curve). In the first case, the differential migration brought
planets into a 3:2 commensurability and in the second case into a 4:3.  
The semi-major axis ratio
of planets librates exactly within the regions of the resonance width marked
in Fig.~\ref{fig5} (right panel)
by  solid, horizontal lines.
Both commensurabilities shown in Fig.~\ref{fig5} are stable for the whole
time of our simulations, namely 
2500 orbits (the unit of time is defined as the orbital period on the
initial orbit of the Super-Earth divided by 2$\pi$). 
For full discussion of these results see  Podlewska \& 
Szuszkiewicz 
(\cite{edytas})
and also Podlewska (this volume).

\section{Conclusions}
Particular attention has been devoted here to the
investigation of the occurrence of resonant configurations in
systems containing low-mass planets.
Studies of
commensurabilities have already been applied successfully to analyse
the motion of pulsar
planets and they proved to be a powerful tool in that context.
They have the  potential to be similarly useful in our search for Earth-like
planets in other systems. 
Our prediction of the occurrence of the commensurabilities in such systems
may turn out to be particularly valuable in the case of the TTV 
(Transit Timing Variation) observations,
because the differences in the time intervals between successive transits,
caused by planet-planet
interactions, are largest near mean-motion resonances
(e.g. Agol {\em et al.\/} \cite{agol}).
Finally, detection of such resonances can also yield useful
information about orbital migration as a process operating during
planet formation.

\section*{Acknowledgements}
This work has been partially supported by MNiSW grant N203~026~32/3831
(2007-2010), the project ASTROSIM-PL and KITP, Santa Barbara, NSF Grant No
PHY99-0794.
E.S. would like to express  her gratitude for support
through a PPARC funded visitor's grant
and for the provision of computer facilities
at the  Astronomy Unit, Queen Mary, University of London.
The simulations reported here were also performed using the
Polish National Cluster of Linux Systems (CLUSTERIX) and the
computational cluster HAL9000 of the Faculty of Mathematics and Physics
at the University of Szczecin.


\end{document}